%% file: main.tex
\documentclass[sigconf, screen]{acmart}
\usepackage{xcolor}
\usepackage{array}
\usepackage{pifont}
\definecolor{commentgreen}{RGB}{9,136,66}

\startPage{1}
\usepackage{algorithm}
\usepackage{algpseudocode}
\usepackage{graphicx} 
\usepackage{booktabs} 
\usepackage{array}
\usepackage{tabularx}
\usepackage{subcaption}
\usepackage{enumitem}

\usepackage{multirow}
\usepackage{makecell}
\usepackage{array}
\usepackage{graphicx}
\usepackage{booktabs}
\usepackage{graphicx}

\copyrightyear{2026}
\acmYear{2026}
\setcopyright{cc}
\setcctype{by}
\acmConference[ICPP '26]{Proceedings of the 55th International Conference on Parallel Processing}{September 28-October 01, 2026}{Singapore, Singapore}
\acmBooktitle{Proceedings of the 55th International Conference on Parallel Processing (ICPP '26), September 28-October 01, 2026, Singapore, Singapore}
\acmDOI{10.1145/3832810.3832821}
\acmISBN{979-8-4007-2657-6/2026/09}

\begin{document}

\title{DB-SpMSpV: Dual-View Blocked Sparse Matrix-Sparse Vector Multiplication for Dynamic GPU Workloads}

\author{Xing Cong}
\affiliation{
  {Beihang University}\\
  \city{Beijing}
  \country{China}
}
\email{congxing@buaa.edu.cn}

\author{Chenhao Xie}
\authornote{Corresponding author}
\affiliation{
  {Beihang University}\\
  \city{Beijing}
  \country{China}
}
\email{xiechenhao@buaa.edu.cn}

\author{Rui Wang}
\affiliation{
  {Beihang University} \\
  \city{Beijing}
  \country{China}
}
\email{wangrui@buaa.edu.cn}

\author{Zhongzhi Luan}
\affiliation{
  {Beihang University}\\
  \city{Beijing}
  \country{China}
}
\email{07680@buaa.edu.cn}

\author{Yi Liu}
\affiliation{
  {Beihang University}\\
  \city{Beijing}
  \country{China}
}
\email{yi.liu@buaa.edu.cn}

\author{Depei Qian}
\affiliation{
  {Beihang University}\\
  \city{Beijing}
  \country{China}
}
\email{depeiq@buaa.edu.cn}

\begin{abstract}
Sparse Matrix-Sparse Vector Multiplication (SpMSpV) is a core primitive in graph traversal, sparse linear algebra, and sparse model inference. Its input vector is often dynamically sparse, so the best GPU execution path depends on both global sparsity and the local vector-block distribution. Existing GPU SpMSpV methods often bind storage layouts, push/pull traversal, and kernels together, making fine-grained adaptation difficult without extra storage or scheduling overhead.

This paper presents DB-SpMSpV, a dual-view blocked SpMSpV framework for dynamic GPU workloads. DB-SpMSpV partitions the matrix into fixed-size 2D blocks, maintains block-level CSR/CSC views at the high level, and reuses a single low-level block payload to support both row-driven pull and column-driven push. At runtime, it selects the global traversal path based on input block sparsity, chooses block microkernels from the local matrix/vector block structure, and uses load balancing, asynchronous prefetching, and hierarchical writeback to reduce irregular memory accesses, writeback conflicts, and load imbalance. We further integrate the framework into DB-BFS and DB-Decoding.

We evaluate DB-SpMSpV on NVIDIA A100 and RTX 4090 using SuiteSparse matrices, symmetric graphs, and three open-source LLMs. Across input sparsities, DB-SpMSpV achieves average speedups of 5.48$\times$--64.34$\times$ over cuSPARSE and 2.36$\times$--14.01$\times$ over TileSpMSpV on A100, with similar gains on RTX 4090. DB-BFS further improves end-to-end graph traversal by 2.66$\times$ over TileBFS on A100 and 3.60$\times$ on RTX 4090 on average, while DB-Decoding accelerates single-token linear layers by up to 4.50$\times$.

\end{abstract}

%% The code below is generated by the tool at http://dl.acm.org/ccs.cfm. CCS




\maketitle 

\protect\input{./sections/introduction}

\protect\input{./sections/motivation}

\protect\input{./sections/method}

\protect\input{./sections/experiment}

\protect\input{./sections/related_work}

\protect\input{./sections/conclusion}

\begin{acks}
We sincerely thank the anonymous reviewers for their insightful suggestions. The work is supported by the Natural Key Research and Develop-ment Program of China (2023YFB3002902).
\end{acks}

% \clearpage 

\bibliographystyle{ACM-Reference-Format}
\bibliography{main}

\end{document}

%% file: sections/introduction.tex
%\vspace{-5pt}

\section{Introduction}

\label{section:1}

Sparse matrix computation is a fundamental technique in computational science \cite{Anzt2017}, graph analytics\cite{davis2019algorithm,kepner2015graphs,sundaram2015graphmat,anderson2016graphpad,yang2022graphblast}, and machine learning systems\cite{fan2025spinfer,ye2025flashinfer}. On GPUs, however, sparse workloads are difficult to accelerate because compact storage and irregular sparsity often lead to scattered memory accesses, load imbalance, and parallel writeback conflicts \cite{greathouse2014efficient,liu2015csr5,maggioni2013adell}. These challenges become more pronounced when the sparsity pattern changes at runtime.

Sparse Matrix-Sparse Vector Multiplication (SpMSpV) captures this dynamic setting. It is a core primitive in Breadth-First Search (BFS)\cite{azad2017work,Ziche2024GPUAcceleratedBF,Niu2025BerryBeesBF}, graph traversal\cite{Sha2019GPUbasedGT}, sparse linear algebra\cite{azad2017work}, and GraphBLAS-like systems\cite{yang2022graphblast,anderson2016graphpad}. It also appears in sparse large language model inference, where sparse weight matrices\cite{Frantar2023SparseGPTML} and sparse activations\cite{Liu2024TrainingFreeAS, cong2026rtlynx} turn some single-token linear transformations into SpMSpV-like operations. Thus, SpMSpV is not only a graph primitive, but also a useful abstraction for workloads that repeatedly multiply a reusable sparse matrix by a changing sparse vector.

\begin{figure}[t!]
    \centering
    \includegraphics[width=1\linewidth]{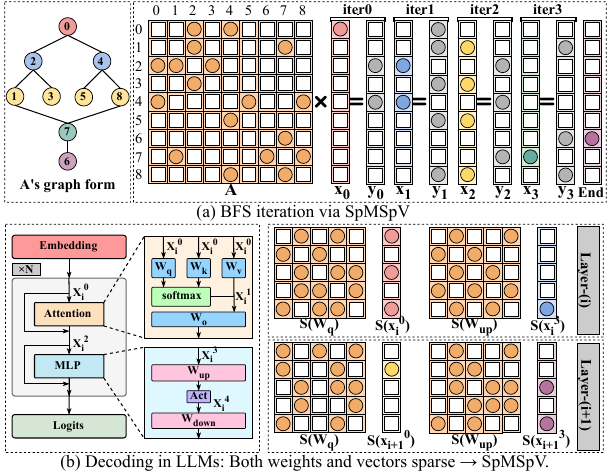}
    \vspace{-20pt}
    \caption{Dynamic SpMSpV in BFS and sparse LLM decoding.}
    \vspace{-20pt}
    \label{fig:bfs-decoding-demo}
\end{figure}

In real workloads, the sparse vector is not fixed. As shown in Fig.\ref{fig:bfs-decoding-demo}, BFS repeatedly applies SpMSpV on a fixed adjacency matrix while its frontier expands, shrinks, and shifts across traversal levels. Sparse LLM decoding similarly reuses weight matrices while activation vectors vary across layers, tokens, and contexts. Although these workloads have different semantics, they require the same systems support: a matrix representation that can be reused across many dynamic vectors, low-overhead vector metadata generation, and execution that adapts as the sparse vector changes.

This requirement makes SpMSpV fundamentally different from SpMV\cite{Niu2021TileSpMVAT, Cong2025CBSpMVADA}. SpMV assumes a dense input vector and usually traverses all matrix nonzeros, whereas SpMSpV\cite{Shah2015SparseMS,Ji2022TileSpMSpVAT,Li2020AdaptiveSO,Li2026VDHAVH} should exploit only the active input positions. As a result, the preferred SpMSpV path is input-dependent: push and pull trade off invalid scans, memory regularity, and writeback conflicts as sparsity changes. In addition, as Fig.\ref{fig:insight-x} shows, a globally sparse BFS frontier contains distinct local regions, making purely global decisions too coarse for efficient computation.

Existing optimizations only partially address this tension. Storage formats such as CSR/CSC\cite{liu2015csr5,azad2017work}, TileSpMSpV\cite{Ji2022TileSpMSpVAT}, and B2SR/bitmap\cite{Niu2025BerryBeesBF} improve either row access, column access, or locality, but are usually aligned with one computation paradigm. Keeping both CSR and CSC supports both push and pull, but nearly doubles storage and complicates switching. GPU kernels such as k-way merge\cite{Li2018MergeBasedPS}, fgSpMSpV\cite{Chen2022fgSpMSpVAF}, VDHA\cite{Li2026VDHAVH}, HAM-SpMSpV\cite{Xu2024HAMSpMSpVAO}, and Tensor Core-based designs\cite{Niu2025BerryBeesBF} reduce specific bottlenecks, yet they are often tied to a particular layout, sparsity range, or application pipeline\cite{Ji2022TileSpMSpVAT,Xu2026CGAAB,Elbek2025BLESTBE}. What is still missing is a coordinated SpMSpV design that supports both push and pull over a shared representation while adapting to global vector sparsity and local block structure.

To this end, we propose DB-SpMSpV, a GPU-based workload that coordinates dual-view blocked storage with two-level runtime adaptation for dynamic SpMSpV. DB-SpMSpV is built on the observation that \textbf{2D blocking decouples inter-block traversal order from intra-block data organization}. At the high level, it maintains block-level CSR and CSC views so that row-block-driven pull and column-block-driven push can share one matrix representation. At the low level, it stores only one copy of the block data and selects the COO, Dense, ForceCSR, ForceCSC, or patched-CSR format based on the intra-block structure. At runtime, DB-SpMSpV selects the traversal path using global nonzero-vector-block ratio, chooses block microkernels using local vector-block masks, and applies load balancing, asynchronous prefetching, warp-level reduction, and hierarchical writeback to improve GPU efficiency.

\begin{figure}[t]
    \centering
    \includegraphics[width=1\linewidth]{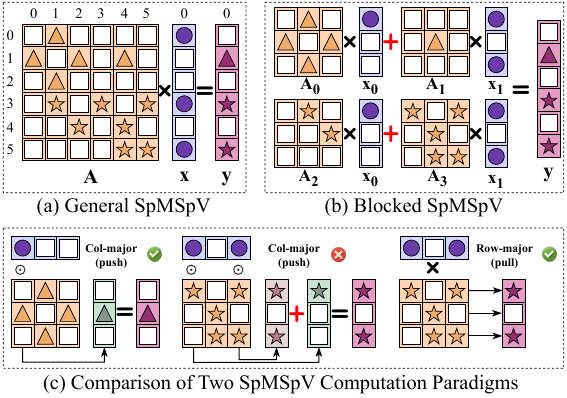}
    \vspace{-20pt}
    \caption{\small{\textbf{Overview of General and Blocked SpMSpV with Push/Pull Computation Paradigms.}}}
    \vspace{-20pt}
    \label{fig:spmspv-demo}
\end{figure}

We further extend DB-SpMSpV to DB-BFS and DB-Decoding. In DB-BFS, dynamic frontiers act as sparse vectors over a fixed graph matrix, enabling push–pull switching without rebuilding the representation. In DB-Decoding, sparse weights are converted once at model loading, while token-dependent activation masks drive adaptive SpMSpV for single-token decoding. Experiments on A100 and RTX 4090 show that DB-SpMSpV outperforms cuSPARSE, CB-SpMV, and TileSpMSpV across input sparsities, and that the same design improves end-to-end BFS and single-token sparse linear layers. The main contributions of this coordinated design are as follows:

% %\vspace{-10pt}
\begin{itemize}[left=0pt]
    \item We propose a dual-view blocked matrix representation that unifies push and pull with low storage overhead.
    \item We design an adaptive DB-SpMSpV framework that selects execution paths using both global block sparsity and local vector-block distribution.
    \item We develop an optimized GPU kernel with path-aware load balancing, asynchronous prefetching, warp-level reduction, and hierarchical writeback.
    \item We extend DB-SpMSpV to DB-BFS and DB-Decoding, showing its effectiveness for graph traversal and sparse LLM decoding.
    \item We perform a comprehensive evaluation on two GPUs across SpMSpV, BFS, and sparse LLM decoding workloads.
\end{itemize}

%% file: sections/motivation.tex
\begin{figure*}
    \centering
    \includegraphics[width=1\linewidth]{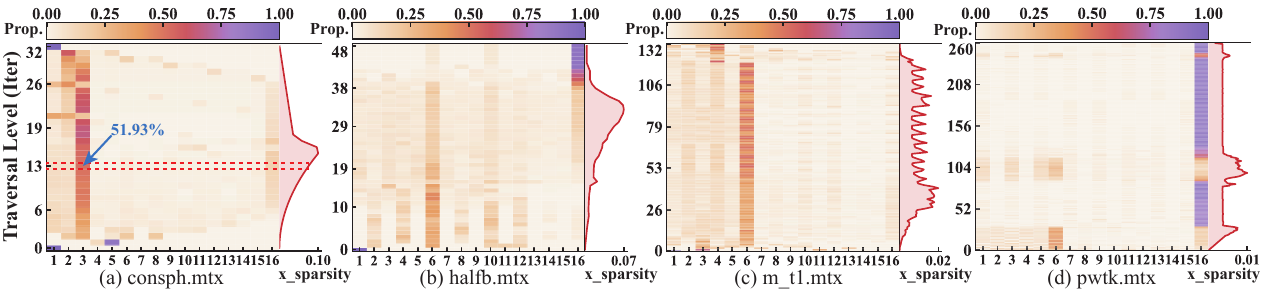}
    \vspace{-15pt}
    \caption{Block-level nonzero distribution of BFS frontiers across iterations. Each frontier is partitioned into 16-element vector blocks. The heatmap shows the proportion of blocks with different numbers of nonzeros (e.g., (3, 13, 0.5193) denotes that vector blocks with 3 nonzeros account for 51.93\% of all nonzero vector blocks), and the red curve shows frontier sparsity $s_f=\mathrm{nnz}(f)/|f|\in[0,1]$.}
    \vspace{-10pt}
    \label{fig:insight-x}
\end{figure*}

\section{Background and Motivation}
\label{section:2}

\subsection{SpMSpV}

Sparse Matrix-Sparse Vector Multiplication (SpMSpV) computes the product of a sparse matrix $A$ and a sparse vector $x$, i.e., $y=A x$, as illustrated in Fig.\ref{fig:spmspv-demo}. Since only part of $x$ is active, the computation can be organized in two opposite directions. Push starts from active input entries and propagates their contributions to output rows, while pull starts from output rows and gathers contributions from the corresponding input entries. These two paradigms expose different memory access patterns, task granularities, and writeback behaviors, making the structure of $x$ central to performance.

\textbf{Push paradigm}: Push is a column-driven approach (Alg.\ref{alg:push-pull-spmspv} Top). For each active entry $j$ in $x$, it traverses column $j$ of $A$ and accumulates $A_{i,j} \times x_j$ into $y_i$. By skipping inactive columns, push can greatly reduce unnecessary memory accesses when $x$ is highly sparse, and it naturally matches CSC-like layouts. Its weakness becomes apparent when many columns are active: different columns may update the same output entries, increasing the number of atomic operations and writeback conflicts. Existing methods such as fgSpMSpV\cite{Chen2022fgSpMSpVAF}, FastSpMSpV\cite{Yang2015FastSM}, and VDHA\cite{Li2026VDHAVH} mitigate these conflicts through hash tables, local aggregation, or decomposition, but writeback pressure remains a major bottleneck as the input becomes denser.

\textbf{Pull paradigm}: Pull is a row-driven approach (Alg.\ref{alg:push-pull-spmspv} Bottom). It traverses each output row $i$ and checks whether the corresponding input entry $x_j$ is active. This direction fits CSR-like layouts, provides more contiguous matrix access, and largely avoids write conflicts because output rows are handled independently. When $x$ is relatively dense, most row scans are useful and pull can provide stable throughput. When $x$ is highly sparse, however, pull still visits many nonzeros whose input entries are zero. Methods such as CSR5\cite{liu2015csr5}, TileSpMV\cite{Niu2021TileSpMVAT}, CB-SpMV\cite{Cong2025CBSpMVADA}, and TileSpMSpV\cite{Ji2022TileSpMSpVAT} improve locality and load balance, but a fixed pull or SpMV-style execution path cannot fully exploit strong input sparsity.

\begin{algorithm}[h]
\caption{Pseudocode of push- and pull-based SpMSpV}
\label{alg:push-pull-spmspv}
\begin{algorithmic}[1]
    \State \textcolor{commentgreen}{ // Push-based SpMSpV (CSC format)}
    \For{$j \in \mathrm{nnz}(x)$ \textbf{in parallel do}} 
        \For{$k = col\_ptr[j]$ to $col\_ptr[j+1]$ do}
            \State $row \gets row\_idx[k]$ \textcolor{commentgreen}{// nonzeros in column $j$}
            \State \textcolor{commentgreen}{// Writeback violation occurred, reducing performance}
            \State $y[row] \gets y[row] + x[j] \times csc\_val[k]$
        \EndFor
    \EndFor
    \State \textcolor{commentgreen}{ // Pull-based SpMSpV (CSR format)}
    \For{$i = 0$ to $m$ \textbf{in parallel do}}
        \State $sum \gets 0$
        \For{$j = row\_ptr[i]$ to $row\_ptr[i+1]$ do}
            \State \textcolor{commentgreen}{// Invalid calculation caused by $x$ being zero}
            \State $sum \gets sum + x[col\_idx[j]] \times csr\_val[j]$
        \EndFor
        \State $y[i] \gets sum$
    \EndFor
\end{algorithmic}
\end{algorithm}

\subsection{Motivation: Need for Two-Level Adaptivity}

\begin{figure}[t!]
    \centering
    \includegraphics[width=1.0\linewidth]{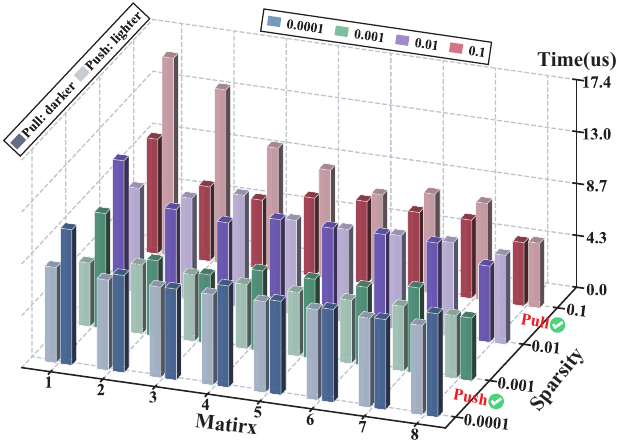}
    \vspace{-15pt}
    \caption{Execution time of CSR-pull and CSC-push SpMSpV on eight matrices under different input sparsities.}
    \vspace{-10pt}
    \label{fig:pull-vs-push}
\end{figure}

To demonstrate this complementarity, we selected eight SuiteSparse matrices and measured the performance of basic CSR-pull and CSC-push implementations across different input vector sparsities, as shown in Fig.\ref{fig:pull-vs-push}. The preferred paradigm shifts as $x$ becomes sparser. At sparsity 0.1, CSR-pull is faster on all eight matrices; for example, on \texttt{matrix-1}, pull takes 9.61us while push takes 16.59us. At sparsity 0.001, push becomes faster on most matrices; for example, on \texttt{matrix-4}, push takes 5.40us, outperforming pull at 6.74us.

The switching behavior, however, is not determined solely by global sparsity. Different matrices have different crossover points: at sparsity 0.01, \texttt{matrix-1} favors push (8.72us vs. 10.85us), while \texttt{matrix-2}, \texttt{matrix-3}, and \texttt{matrix-6} favor pull by contrast. Push performance is also not always monotonic with sparsity; on \texttt{matrix-1}, push time increases from 5.33us at sparsity 0.001 to 8.01us at sparsity 0.0001. These results indicate that the optimal path depends not only on the number of active inputs, but also on how those inputs interact with the matrix structure.

\begin{figure*}[t!]
    \centering
    \includegraphics[width=1\linewidth]{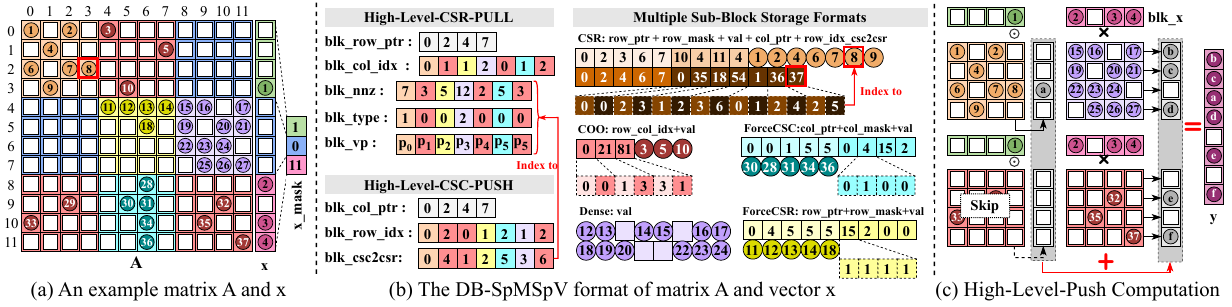}
    \vspace{-15pt}
    \caption{Storage format and execution flow of DB-SpMSpV. (a) A $12\times12$ example matrix with different block formats and vector-block metadata, with block size 4; (b) high-level dual-view indices and low-level block data. Here, \texttt{xxx\_mask} denotes a bit mask indicating nonzero positions; for example, the first row\_mask value of \texttt{10} in CSR corresponds to \texttt{0b1010}, indicating nonzeros in columns 1 and 3; (c) block-level computation flow under the high-level push path for a sparse input vector.}
    \vspace{-5pt}
    \label{fig:DB-format}
\end{figure*}

\begin{figure}[t!]
    \centering
    \includegraphics[width=1\linewidth]{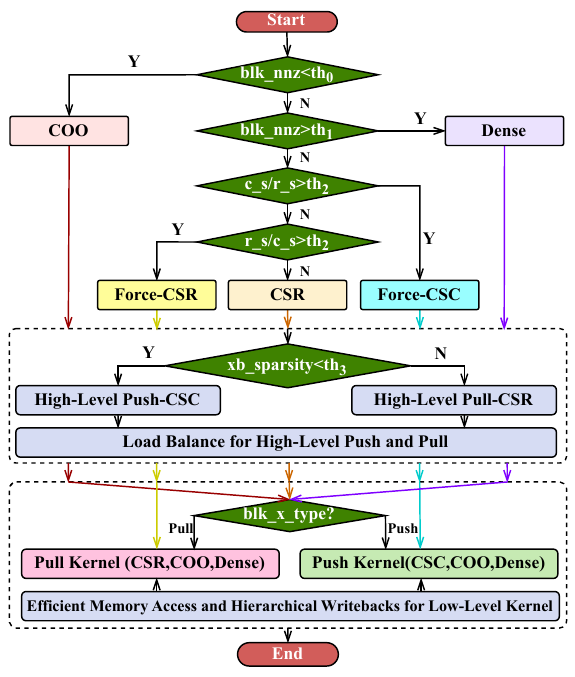}
    \vspace{-20pt}
    \caption{Overview of DB-SpMSpV.}
    \vspace{-15pt}
    \label{fig:overview}
\end{figure}

Meanwhile, we observe that matrices in real applications exhibit strong local heterogeneity. We select four matrices(in Tab.\ref{tab:bfs-representative}) with distinct structural properties---\texttt{pwtk} and \texttt{consph} are moderately dense finite-element matrices, \texttt{m\_t1} has a strongly banded structure with locally dense blocks, and \texttt{halfb} exhibits irregular sparsity---to cover the spectrum of frontier evolution patterns encountered in practice. Using vertex 0 as the BFS source, we split each frontier into 16-element vector blocks and profile the number of nonzeros inside each block, as shown in Fig.\ref{fig:insight-x}. The frontiers are highly nonuniform across both matrices and iterations. For example, in iteration 13 of \texttt{consph}, blocks with 3 nonzeros account for 51.93\%, while full blocks still account for 16.05\%. In \texttt{pwtk}, full blocks exceed 80\% in most iterations, meaning that a globally sparse frontier can still contain locally dense regions. Similar patterns appear in \texttt{halfb} and \texttt{m\_t1}.

These observations lead to the main challenge addressed in this paper: \textbf{SpMSpV needs adaptivity at two levels}. At the global level, it should choose whether to traverse active input or output. At the local level, it should determine how the matrix is processed based on its structure and the sparse vector. A single push or pull path is therefore insufficient, while maintaining fully duplicated row- and column-oriented formats is too expensive. This motivates a unified blocked representation that supports block-level push/pull collaboration with low overhead. As Fig.\ref{fig:spmspv-demo}(b) shows, 2D blocking provides a middle ground by exposing coarse-grained matrix blocks traversal and individual intra-blocks computing model selection, transferring the key challenges to optimal intra-block designing and inter-block scheduling.

%% file: sections/method.tex
\section{DB-SpMSpV}

\label{section:3}

\subsection{Overview}

To achieve our goal, we propose DB-SpMSpV, a dual-view blocked SpMSpV framework for dynamic GPU workloads, tailored to the characteristics of both matrices $A$ and $x$. DB-SpMSpV uses fixed-size 2D blocks as the common unit for storage, scheduling, and local computation. As shown in Fig.\ref{fig:overview}, it first converts the matrix into 16$\times$16 blocks with high-level row/column traversal views and density/skew-aware low-level formats. At runtime, the input vector is partitioned into aligned 16-element blocks to generate masks and global statistics that drive two-level adaptivity: selecting the high-level push/pull traversal path and choosing block microkernels based on the local matrix/vector structure. The same adaptive SpMSpV engine is then reused by DB-BFS and DB-Decoding.

\subsection{Dual-view sparse storage format}

\subsubsection{\textbf{Sparse matrix storage format}}

DB-SpMSpV partitions the input matrix into uniform 16$\times$16 blocks. This granularity matches warp-level execution and provides enough local structure for efficient block microkernels, while avoiding excessive scheduling overhead from very small blocks or insufficient parallelism from very large blocks. The representation is organized in two layers. The high-level layer locates nonzero blocks and provides traversal views, whereas the low-level layer stores the internal structure and values of each block.

The high-level layer maintains both block-level CSR and block-level CSC views, as shown in Fig.\ref{fig:DB-format}(b). The block-level CSR view, represented by \texttt{global\_block\_csr\_row\_ptr} and \texttt{global\_block\_csr\_col\_idx}, enumerates nonzero blocks from output row blocks and serves as the entry point of the pull path. The block-level CSC view, represented by \texttt{global\_block\_csc\_col\_ptr} and \texttt{global\_block\_csc\_row\_idx}, enumerates nonzero blocks from active input column blocks and serves as the entry point of the push path. The block payload is stored only once in a primary order. A mapping array, \texttt{global\_block\_csc2csr}, translates each CSC-view position to the corresponding primary block ID, allowing both views to reuse the same \texttt{blocks\_row}, \texttt{blocks\_col}, \texttt{blocks\_nnz}, \texttt{blocks\_type}, and \texttt{blocks\_vp} metadata. This design preserves both traversal directions without storing a duplicate CSR+CSC.

At the low level, DB-SpMSpV selects compact block formats by density and directional skew. Blocks with fewer than $th_0$ nonzeros use packed COO, blocks with more than $th_1$ nonzeros use Dense storage, and the remaining blocks are classified by row/column imbalance. For axis $a \in \{r,c\}$, let $B_a$ be the number of rows or columns and $n_l^{(a)}$ be the nonzero count of the $l$-th row or column. We define the axis skew as:
$$
\text{skew}_a(A_b)=\frac{\max_l n_l^{(a)}}{\mathrm{nnz}(A_b)/B_a+\epsilon},
$$
where $\epsilon$ avoids division by zero. Here, $\text{skew}_r$ and $\text{skew}_c$ correspond to row and column skew, respectively. If $\frac{\text{skew}_c}{\text{skew}_r} > th_2$, the block is marked as ForceCSC; if $\frac{\text{skew}_r}{\text{skew}_c} > th_2$, it is marked as ForceCSR. Moderately sparse blocks without strong skew use patched CSR, which adds \texttt{col\_ptr} and packs 4-bit \texttt{row\_idx} with 4-bit \texttt{csc2csr} into \texttt{row\_idx\_csc2csr}, allowing push to reuse the same value array without a separate CSC payload. Following CB-SpMV\cite{Cong2025CBSpMVADA}, we set $th_0=32$, $th_1=128$, and $th_2=2.0$, and store block metadata contiguously for cache locality.

%%\vspace{-10pt}
\begin{algorithm}[h]
\caption{Pseudocode of warp-level DB-SpMSpV for a ForceCSR block in the pull paradigm}
\label{alg:db-force-csr}
\small
\begin{algorithmic}[1]
    \For{$t_i=0$ \textbf{to} $31$ \textbf{in parallel}}
        \State $sum \gets 0$, $nz\_idx \gets 0$
        \State \textcolor{commentgreen}{// Two threads interleave to process one row}
        \State $row\_idx \gets t_i \gg 1$
        \State $sub\_lane\_idx \gets t_i \,\&\, 1$
        \State $active \gets row\_mask[row\_idx] \,\&\, blk\_x\_mask$
        \While{$active \neq 0$}
            \State $col\_idx \gets \text{\_\_ffs}(active)-1$
            \If{$(nz\_idx \,\&\, 1) = sub\_lane\_idx$}
                % \State $offset \gets \text{\_\_popc}(row\_mask[row\_idx] \,\&\, ((1 \ll col\_idx)-1))$
                \State $prefix\_mask \gets \text{mask of columns before } col\_idx$
                \State $offset \gets \text{\_\_popc}(row\_mask[row\_idx] \,\&\, prefix\_mask)$
                \State $p \gets row\_ptr[row\_idx] + offset$
                \State $sum \gets sum + val[p] \times blk\_x[col\_idx]$
            \EndIf
            \State $active \gets active \,\&\, (active - 1)$
            \State $nz\_idx \gets nz\_idx + 1$
        \EndWhile
        \State \textcolor{commentgreen}{// Consolidated results}
        \State $sum \gets sum + \text{\_\_shfl\_xor\_sync}(\text{0xffffffff}, sum, 1)$
        \If{$sub\_lane\_idx = 0$}
            \State $s\_y[row\_idx] \gets sum$
        \EndIf
    \EndFor
\end{algorithmic}
\end{algorithm}
%%\vspace{-10pt}

\subsubsection{\textbf{Sparse vector storage format}}

As shown in Fig.\ref{fig:DB-format}(a), the input vector is also partitioned into 16-element vector blocks, aligned with the column dimension of matrix blocks. Besides the original vector $x$, each vector block stores \texttt{blk\_x\_mask}, a 16-bit mask that records intra-block nonzero positions. All-zero blocks have a zero mask and can be skipped immediately. During computation, \texttt{\_\_popc()} counts the number of active entries in a vector block: fewer active entries favor push, while denser vector blocks favor pull. This vector representation supplies the local information needed for block-level microkernel selection, while the global nonzero-block statistics guide high-level traversal selection.

\subsection{DB-SpMSpV Algorithm}

The dual-view format provides the structural basis for both traversal directions, but performance depends on how the runtime uses it. DB-SpMSpV performs two-level adaptive execution. First, it selects a high-level traversal path according to the ratio of nonzero vector blocks. Second, for each visited matrix block, it selects a low-level microkernel according to the matrix block type and \texttt{blk\_x\_mask}. This separates global traversal efficiency from local computation and writeback efficiency. To sustain GPU throughput under irregular sparsity, DB-SpMSpV further incorporates path-aware load balancing, asynchronous data loading, and hierarchical writeback.

\subsubsection{\textbf{Dual-Path Adaptive Execution}}

DB-SpMSpV first scans vector blocks to count active entries and nonempty vector blocks, then computes the nonempty-block sparsity $s_b=N_{\mathrm{nzblk}}/N_{\mathrm{blk}}\in[0,1]$ (\texttt{xb\_sparsity}). If \texttt{xb\_sparsity} is below $th_3=0.01$, the high-level path uses push and enumerates only matrix blocks associated with active input column blocks. Otherwise, the high-level path uses pull and organizes work by output row blocks, reducing writeback conflicts when many vector blocks are active. After the high-level path is fixed, DB-SpMSpV still selects a push or pull microkernel for each block based on \texttt{blk\_x\_mask} and \texttt{blocks\_type}. High-level decisions, therefore, control which matrix blocks are visited, while low-level decisions control how each visited block is computed.

%\vspace{-8pt}
\begin{algorithm}[h]
\caption{Pseudocode of warp-level DB-SpMSpV for a ForceCSC block in the push paradigm}
\label{alg:db-force-csc}
\small
\begin{algorithmic}[1]
    \For{$t_i=0$ \textbf{to} $31$ \textbf{in parallel}}
        \State \textcolor{commentgreen}{// Two half-warps process different column ranges for the rows}
        \State $row\_idx \gets t_i \,\&\, 15$, \quad $part \gets t_i \gg 4$
        \State $active \gets blk\_x\_mask$, $sum \gets 0$
        \State $active \gets active \,\&\, (\text{0x00ff} \text{ if } part = 0 \text{ else } \text{0xff00})$
        \While{$active \neq 0$}
            \State $col\_idx \gets \text{\_\_ffs}(active)-1$
            \State $mask \gets col\_mask[col\_idx]$
            \If{$((mask \gg row\_idx) \,\&\, 1) \neq 0$}
                \State $offset \gets \text{\_\_popc}(mask \,\&\, ((1 \ll row\_idx)-1))$
                \State $p \gets col\_ptr[col\_idx] + offset$
                \State $sum \gets sum + val[p] \times blk\_x[col\_idx] $
            \EndIf
            \State $active \gets active \,\&\, (active - 1)$
        \EndWhile
        \State $sum \gets sum + \text{\_\_shfl\_xor\_sync}(\text{0xffffffff}, sum, 16)$
        \If{$t_i < 16$}
            \State $s\_y[row\_idx] \gets sum$
        \EndIf
    \EndFor
\end{algorithmic}
\end{algorithm}
%\vspace{-10pt}

\begin{figure*}[t!]
    \centering
    \includegraphics[width=1\linewidth]{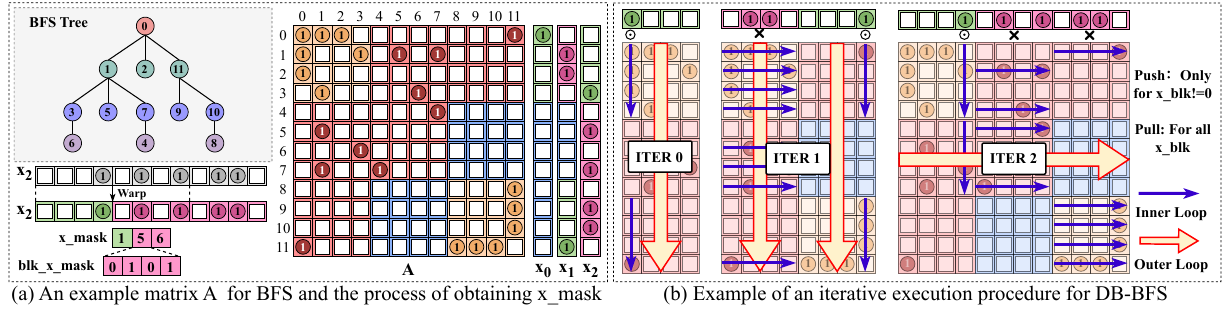}
    \vspace{-15pt}
    \caption{DB-BFS execution with frontier mask generation and adaptive push/pull traversal. (a) An example BFS tree, its adjacency matrix, and the generation of \texttt{x\_mask} from the current frontier. (b) Iterative DB-BFS execution, where push is applied only to active vector blocks and pull scans all relevant vector blocks, enabling path switching across frontier evolution.}
    \vspace{-10pt}
    \label{fig:db-bfs}
\end{figure*}

\subsubsection{\textbf{Path-Aware Load Balancing}}

Push and pull require different load-balancing strategies. Push may expose too few active column blocks or highly uneven column-block degrees, so DB-SpMSpV flattens work into matrix-block-level tasks: each warp processes one matrix block and fetches tasks in a grid-stride manner, distributing long column blocks across SMs. Pull may create long row-block tasks with uneven format complexity, so DB-SpMSpV splits row blocks into load-weighted segments while preserving local writeback aggregation. Each pull thread block contains four warps and processes a row-block segment rather than a full row block. Segment weights follow the estimated format cost: Dense is 4, ForceCSR/ForceCSC/CSR is 2, and COO is 1; a new segment is created once the accumulated weight reaches 16.

\subsubsection{\textbf{Memory-Latency Hiding and Hierarchical Writeback}}

On GPUs, SpMSpV is usually limited by irregular memory access and writeback synchronization rather than floating-point throughput. DB-SpMSpV therefore uses the fixed block granularity to coordinate data loading, local reduction, and writeback.

%\vspace{-8pt}
\begin{algorithm}[h]
\caption{Pseudocode of warp-level DB-SpMSpV for a Dense block in the pull paradigm}
\label{alg:db-dense}
\small
\begin{algorithmic}[1]
    \For{$t_i=0$ \textbf{to} $31$ \textbf{in parallel}}
        \State $half \gets t_i \gg 4$
        \State $col\_idx \gets t_i \,\&\, 15$
        \State $reduce\_mask \gets (\text{0x0000ffff} \text{ if } half=0 \text{ else } \text{0xffff0000})$
        \For{$row\_base=0$ \textbf{to} $14$ \textbf{step} $2$}
            \State $row\_idx \gets row\_base + half$
            \State \textcolor{commentgreen}{// Coalesced global-memory load}
            \State $sum \gets val[row\_idx \times 16 + col\_idx] \times blk\_x[col\_idx]$
            \State \textbf{for } $OFS \in \{8,4,2,1\}$ \textbf{ do }  
            \State $sum \gets sum +\text{\_\_shfl\_down\_sync}(reduce\_mask, sum, OFS)$
            \If{$col\_idx = 0$}
                \State $s\_y[row\_idx] \gets sum$
            \EndIf
        \EndFor
    \EndFor
\end{algorithmic}
\end{algorithm}
%\vspace{-10pt}

For data loading, push caches the active vector block in shared memory so consecutive tasks from the same column block can reuse it. Pull uses double-buffered shared memory and overlaps prefetching with current computation through \texttt{\_\_pipeline\_memcpy\_async} and \texttt{\_\_pipeline\_wait\_prior}.

During microkernel execution, DB-SpMSpV performs local reduction before global writeback. COO blocks scan packed coordinates and accumulate in shared memory. ForceCSR uses pull with two interleaved threads per row and register-level reduction, as shown in Alg.\ref{alg:db-force-csr}. ForceCSC uses push by splitting a warp into two half-warps over different column ranges, as shown in Alg.\ref{alg:db-force-csc}; row partial sums are merged within the warp before shared-memory writeback. 

For regular CSR blocks, DB-SpMSpV chooses the local execution paradigm according to the number of active elements in \texttt{blk\_x\_mask}. If the vector block is relatively dense, it uses a ForceCSR-like pull path to reduce by rows and lower write conflicts. If the vector block is sparse, it uses push: for each active column $j$, the microkernel starts from \texttt{col\_ptr[$j$]}, reads packed bytes from \texttt{row\_idx\_csc2csr}---where the upper 4 bits encode the local row index and the lower 4 bits encode the offset into \texttt{val[]}---to recover the row and value positions without scanning irrelevant CSR rows. Dense blocks are naturally efficient under pull because values are stored in row-major order, as shown in Alg.\ref{alg:db-dense}. Under push, column-wise access to this row-major layout becomes strided; thus, all 32 warp threads cooperatively load the $16\times16$ block row-by-row into shared memory (each thread loading one element per step), reorganizing it into a column-accessible layout so that values along each active column can be read contiguously during accumulation.

\begin{figure*}[t!]
    \centering
    \includegraphics[width=1\linewidth]{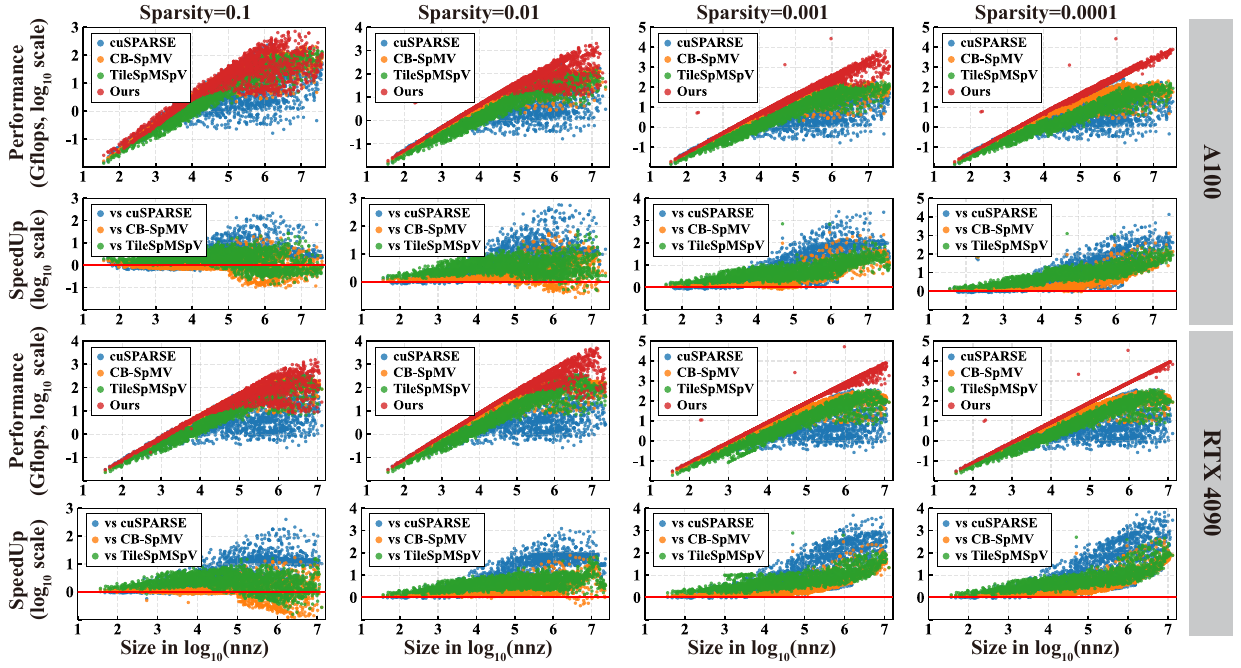}
    \vspace{-20pt}
    \caption{Performance and Speedup of SpMSpV across input sparsities on A100 and RTX 4090.}
    \vspace{-5pt}
    \label{fig:spmspv-somparision}
\end{figure*}

\section{Application-Level Integration}

\label{section:4}

\subsection{Efficient Construction of Vector Metadata}

End-to-end performance also depends on how quickly the dynamic input vector is converted into DB-SpMSpV metadata. In BFS, the frontier can be represented directly as a set of masks. In sparse decoding, activation sparsity must be encoded at runtime. DB-SpMSpV fuses \texttt{blk\_x\_mask} generation and nonzero-block counting on the GPU, and connects this metadata construction to SpMSpV through \texttt{Cooperative Groups} grid-level synchronization. Each warp reads 32 elements, uses \texttt{\_\_ballot\_sync} to obtain a 32-bit nonzero pattern, and splits it into two 16-bit masks. Nonzero blocks are counted hierarchically: each warp checks its assigned vector blocks, thread blocks aggregate locally, and only one global \texttt{atomicAdd} is issued per block. This avoids CPU involvement and host-device transfers, allowing dynamic inputs to be generated and consumed in place on the GPU. The grid-level barrier introduces no additional kernel launch, and its synchronization cost is amortized over the full SpMSpV execution, keeping the online metadata overhead negligible.

\subsection{DB-BFS}

As shown in Fig.\ref{fig:db-bfs}, DB-BFS offline converts the graph adjacency matrix into the DB format and treats each frontier as the sparse input vector of SpMSpV. Because push and pull share the same DB matrix representation, DB-BFS can switch traversal paths across BFS levels without rebuilding the graph; it only updates the frontier masks and vector-block statistics. Sparse frontiers favor push, which skips invalid edge accesses, while expanded frontiers favor pull, which reduces writeback conflicts and improves contiguous memory access. DB-BFS also specializes in microkernels for Boolean propagation. As shown in Alg.\ref{alg:db-bfs-csc}, the kernel only checks reachability and generates the next-frontier mask, thereby avoiding full-precision multiply-adds.

%\vspace{-10pt}
\begin{algorithm}[h]
\caption{Pseudocode of warp-level DB-BFS for a ForceCSC block in the Boolean push paradigm}
\label{alg:db-bfs-csc}
\small
\begin{algorithmic}[1]
    \For{$t_i=0$ \textbf{to} $31$ \textbf{in parallel}}
        \State $found \gets 0$
        \State $row\_idx \gets t_i \,\&\, 15$, \quad $part \gets t_i \gg 4$
        \State $active \gets frontier\_mask \,\&\, (\text{0x00ff} \text{ if } part=0 \text{ else } \text{0xff00})$
        \While{$active \neq 0$ \textbf{and} $found = 0$}
            \State $col\_idx \gets \text{\_\_ffs}(active)-1$
            \State $mask \gets col\_mask[col\_idx]$
            \If{$((mask \gg row\_idx) \,\&\, 1) \neq 0$}
                \State $found \gets 1$
            \EndIf
            \State $active \gets active \,\&\, (active - 1)$
        \EndWhile
        \State \textcolor{commentgreen}{// Merge results from two half-warps}
        \State $flag \gets found \,\lor\, \text{\_\_shfl\_xor\_sync}(\text{0xffffffff}, found, 16)$
        \State \textcolor{commentgreen}{// Remove visited nodes}
        \State $flag \gets flag \,\&\, \neg((visited\_mask \gg row\_idx) \,\&\, 1)$
        \State \textcolor{commentgreen}{// Generate next frontier mask}
        \State $next\_mask \gets \text{\_\_ballot\_sync}(\text{0x0000ffff}, flag \neq 0)$
    \EndFor
    \State \Return $next\_mask$
\end{algorithmic}
\end{algorithm}
%\vspace{-10pt}

\subsection{DB-Decoding}

In DB-Decoding, sparse weight matrices are converted to the DB format during model loading, while runtime activations serve as dynamic sparse input vectors. After each token's activation mask is generated on the GPU, DB-SpMSpV selects execution paths based on the nonzero-block ratio and the intra-block activation distribution. This avoids maintaining multiple weight layouts for different activation sparsities and provides stable sparse-decoding performance across tokens and layers.

%% file: sections/experiment.tex
\begin{figure*}[t!]
    \centering
    \includegraphics[width=1\linewidth]{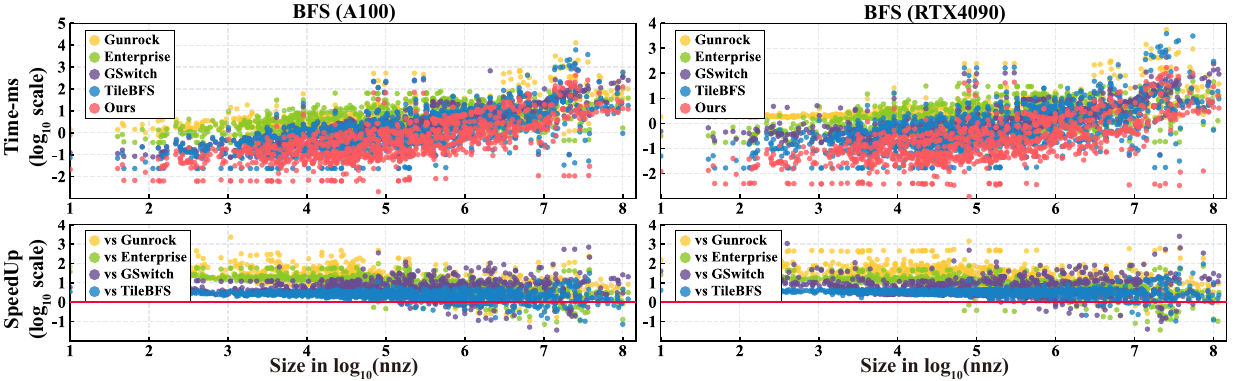}
    \vspace{-10pt}
    \caption{End-to-end BFS performance on A100 and RTX 4090.}
    \vspace{-5pt}
    \label{fig:bfs-comparision}
\end{figure*}

\section{Evaluation}

\label{section:5}

\subsection{Experimental Setup}

We evaluate DB-SpMSpV at both the kernel and application levels on NVIDIA A100 and RTX 4090 GPUs, using driver version 550.135 and CUDA 12.4. Tab.~\ref{tab:environment} summarizes the platforms and baselines. All baselines are open-source state-of-the-art implementations. The SpMSpV kernel evaluation uses 2,738 SuiteSparse matrices from a wide range of domains, including structured grids, social networks, and finite-element problems. BFS evaluation uses over 1,000 symmetric unweighted graphs, and DB-Decoding uses three open-source LLMs under batch-size-1 single-token decoding.

%%\vspace{-10pt}
\begin{table}[h]
\centering
\caption{Algorithms and Machine Specifications.}
\vspace{-5pt}
\label{tab:environment}
\small
\setlength{\tabcolsep}{3pt}
\renewcommand{\arraystretch}{1.15}
\begin{tabularx}{\linewidth}{|c|>{\centering\arraybackslash}p{0.40\linewidth}|X|}
\hline
 & \textbf{Algorithm} & \centering\arraybackslash\textbf{GPU Specification} \\
\hline
\multirow{4}{*}{\textbf{SpMSpV}}
& (1) cuSPARSE v12.4 BSR
& \multirow{7}{=}{%
\par
(1) NVIDIA A100 (Ampere), 6,912 CUDA cores @ 1,410 MHz, 40 GB, B/W 1.56 TB/s. \par
%%\vspace{0.25em}
(2) NVIDIA RTX 4090 (Ada Lovelace), 16,384 CUDA cores @ 2,520 MHz, 24 GB, B/W 1.01 TB/s.
} \\
\cline{2-2}
& (2) CB-SpMV\cite{Cong2025CBSpMVADA} & \\
\cline{2-2}
& (3) TileSpMSpV\cite{Ji2022TileSpMSpVAT} & \\
\cline{2-2}
& (4) \textbf{DB-SpMSpV} (Ours) & \\
\cline{1-2}
\multirow{5}{*}{\textbf{BFS}}
& (1) Gunrock\cite{Wang2017Gunrock} & \\
\cline{2-2}
& (2) Enterprise\cite{Liu2015EnterpriseBG} & \\
\cline{2-2}
& (3) GSwitch\cite{Meng2019APB} & \\
\cline{2-2}
& (4) TileBFS\cite{Ji2022TileSpMSpVAT} & \\
\cline{2-2}
& (5) \textbf{DB-BFS} (Ours) & \\
\cline{1-2}
\multirow{2}{*}{\textbf{LLM-Decoding}}
& (1) PyTorch v2.11 & \\
\cline{2-2}
& (2) \textbf{DB-Decoding} (Ours) & \\
\cline{2-2}
\hline
\end{tabularx}
\end{table}
%%\vspace{-10pt}

%%\vspace{-5pt}
\subsection{Performance of SpMSpV}

For each matrix, we generate random input vectors at different sparsity levels with a fixed seed of 42. We compare DB-SpMSpV with general-purpose cuSPARSE and specialized CB-SpMV and TileSpMSpV; for fairness, cuSPARSE BSR uses the same block size as DB-SpMSpV. All kernels are warmed up and executed 100 times, and performance is reported in GFLOPS using the SpMV floating-point operation count. Fig.\ref{fig:spmspv-somparision} shows that DB-SpMSpV achieves the best performance on both GPUs for most matrices, with a larger advantage as input sparsity decreases from 0.1 to 0.0001. On A100, DB-SpMSpV obtains average speedups of 5.48$\times$–64.34$\times$, 1.20$\times$–8.29$\times$, and 2.36$\times$–14.01$\times$ over cuSPARSE, CB-SpMV, and TileSpMSpV, respectively. On RTX 4090, the corresponding speedups are 8.86$\times$–128.46$\times$, 1.29$\times$–6.98$\times$, and 2.72$\times$–12.04$\times$. The increasing advantage at lower input sparsities is consistent with DB-SpMSpV's vector-block masks and adaptive push/pull execution: inactive regions can be skipped when the vector is sparse, while denser vector blocks can use pull-style computation to improve memory regularity and reduce writeback conflicts.

%%\vspace{-5pt}
\subsection{Application Performance}

\begin{table*}[t]
\centering
\caption{Performance comparison of different linear projections. Time is reported in us, with speedup shown in parentheses.}
\vspace{-5pt}
\label{tab:decoding-comparision}
\small
\setlength{\tabcolsep}{3.2pt}
\renewcommand{\arraystretch}{1.05}
\begin{tabular}{c|cc|cc|cc|cc|cc|cc}
\toprule
\multirow{2}{*}{Model}
& \multicolumn{2}{c|}{Q}
& \multicolumn{2}{c|}{K}
& \multicolumn{2}{c|}{V}
& \multicolumn{2}{c|}{Out}
& \multicolumn{2}{c|}{Up}
& \multicolumn{2}{c}{Down} \\
\cmidrule(lr){2-3}
\cmidrule(lr){4-5}
\cmidrule(lr){6-7}
\cmidrule(lr){8-9}
\cmidrule(lr){10-11}
\cmidrule(lr){12-13}
& PyTorch & Ours
& PyTorch & Ours
& PyTorch & Ours
& PyTorch & Ours
& PyTorch & Ours
& PyTorch & Ours \\
\midrule
Llama-3.2-1B
& 6.53 & 2.69 (2.43$\times$)
& 6.08 & 2.46 (2.47$\times$)
& 6.08 & 2.46 (2.47$\times$)
& 6.52 & 2.64 (2.47$\times$)
& 14.96 & 4.38 (3.42$\times$)
& 13.12 & 4.26 (3.08$\times$) \\

Qwen2.5-1.5B
& 6.02 & 2.56 (2.35$\times$)
& 6.00 & 2.44 (2.46$\times$)
& 6.00 & 2.44 (2.46$\times$)
& 6.03 & 2.54 (2.37$\times$)
& 13.34 & 4.08 (3.27$\times$)
& 19.24 & 4.27 (4.50$\times$) \\

Qwen3-0.6B
& 6.04 & 2.54 (2.38$\times$)
& 6.13 & 2.45 (2.50$\times$)
& 6.00 & 2.46 (2.44$\times$)
& 6.05 & 2.52 (2.40$\times$)
& 6.14 & 2.60 (2.36$\times$)
& 7.03 & 2.79 (2.52$\times$) \\
\bottomrule
\end{tabular}
\end{table*}

\begin{table*}[t]
\centering
\caption{Performance comparison of BFS on representative matrices. Time is reported in ms, and speedup is shown in parentheses.}
\vspace{-5pt}
\label{tab:bfs-representative}
\small
\setlength{\tabcolsep}{3.0pt}
\renewcommand{\arraystretch}{1.08}

\newcommand{\ccell}[1]{\raisebox{-0.5\height}{#1}}
\newcommand{\mtximg}[1]{\ccell{\includegraphics[width=0.085\textwidth,height=0.050\textwidth,keepaspectratio]{mtx/#1.png}}}
\newcommand{\vecimg}[1]{\ccell{\includegraphics[width=0.085\textwidth,height=0.035\textwidth,keepaspectratio]{vector/#1.png}}}

\resizebox{\textwidth}{!}{
\begin{tabular}{c|cccccccc}
\toprule
Metric
& pwtk
& pkustk14
& audikw\_1
& crankseg\_2
& halfb
&msdoor
& consph
& m\_t1 \\
\midrule

\ccell{\makecell[c]{Thumbnail}}
& \mtximg{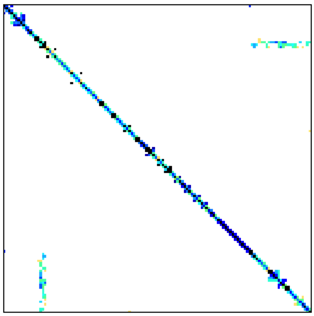}
& \mtximg{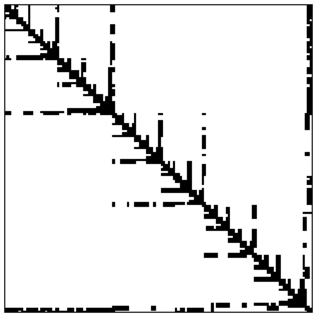}
& \mtximg{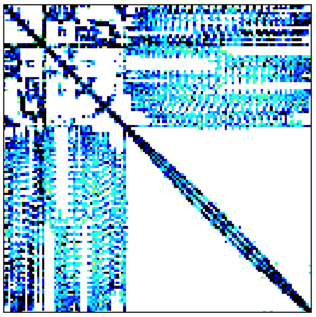}
& \mtximg{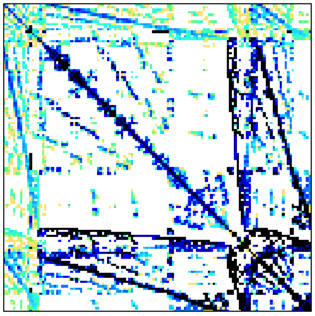}
& \mtximg{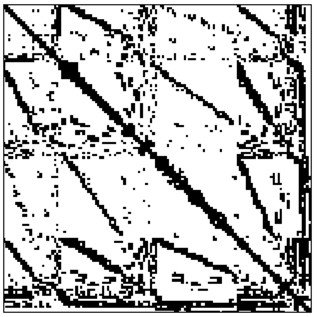}
& \mtximg{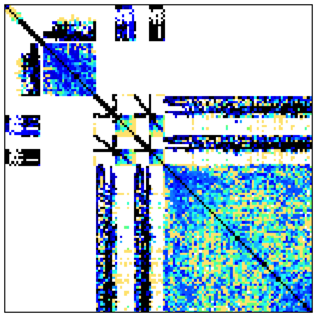}
& \mtximg{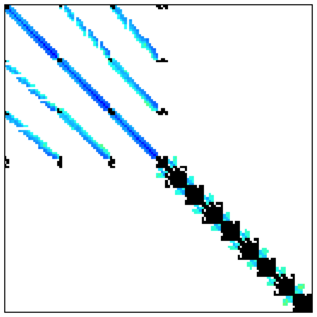}
& \mtximg{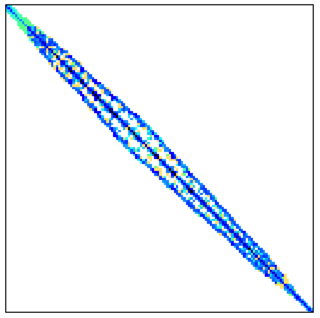} \\
\midrule

\makecell[c]{Row (Column)\\NNZs}
& \makecell[c]{217,918\\11,524,432}
& \makecell[c]{151,926\\14,836,504}
& \makecell[c]{943,695\\77,651,847}
& \makecell[c]{22,283\\24,669,643}
& \makecell[c]{224,617\\12,387,821}
& \makecell[c]{415,863\\19,173,163}
& \makecell[c]{83,334\\6,010,480}
& \makecell[c]{97,578\\9,753,570} \\
\midrule

\ccell{\makecell[c]{Frontier Sparsity}}
& \vecimg{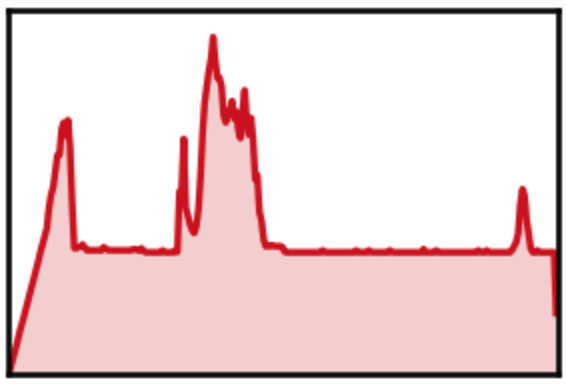}
& \vecimg{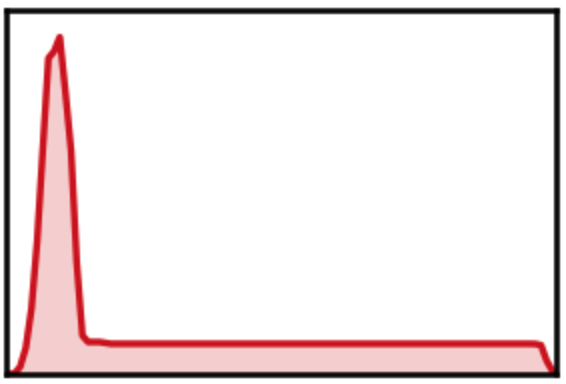}
& \vecimg{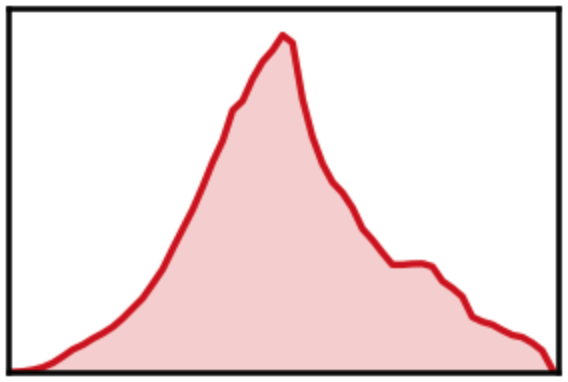}
& \vecimg{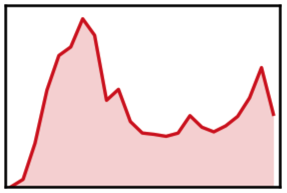}
& \vecimg{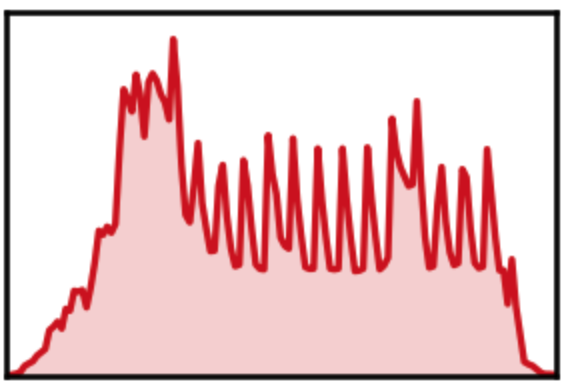}
& \vecimg{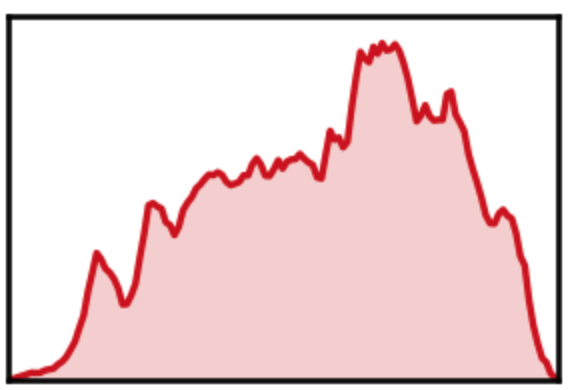}
& \vecimg{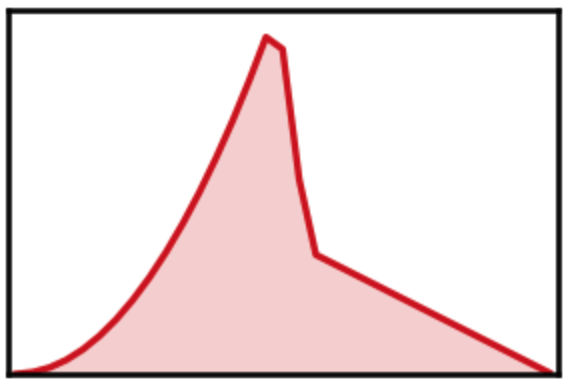}
& \vecimg{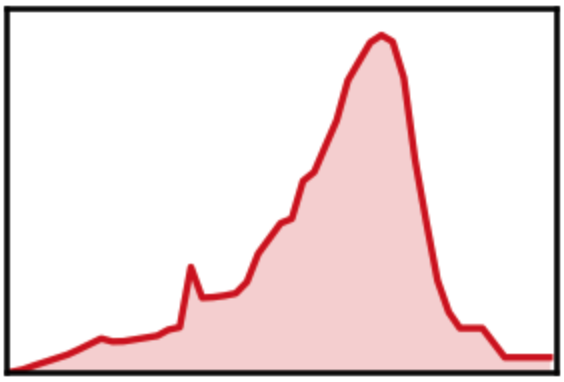} \\
\midrule

Gunrock
& 10.87 (4.30$\times$)
& 6.317 (5.31$\times$)
& 12.49 (4.04$\times$)
& 4.814 (9.47$\times$)
& 6.577 (5.21$\times$)
& 6.45 (4.01$\times$)
& 3.094 (8.92$\times$)
& 4.216 (8.58$\times$) \\

Enterprise
& 30.856 (12.22$\times$)
& 6.292 (5.29$\times$)
& 4.26 (1.38$\times$)
& 1.574 (3.10$\times$)
& 8.354 (6.62$\times$)
& 8.527 (5.30$\times$)
& 2.10 (6.05$\times$)
& 3.013 (6.13$\times$) \\

GSwitch
& 10.299 (4.08$\times$)
& 12.809 (10.76$\times$)
& 66.33 (21.43$\times$)
& 7.905 (15.56$\times$)
& 5.849 (4.63$\times$)
& 5.443 (3.38$\times$)
& 1.453 (4.19$\times$)
& 7.005 (14.25$\times$) \\

TileBFS
& 8.758 (3.47$\times$)
& 3.119 (2.62$\times$)
& 3.490 (1.13$\times$)
& 0.642 (1.26$\times$)
& 3.715 (2.94$\times$)
& 4.261 (2.65$\times$)
& 0.889 (2.56$\times$)
& 1.563 (3.18$\times$) \\

DB-BFS-Pull
& 3.434 (1.36$\times$)
& 1.908 (1.60$\times$)
& 4.077 (1.32$\times$)
& 1.027 (2.02$\times$)
& 3.329 (2.64$\times$)
& 3.617 (2.25$\times$)
& 0.921 (2.65$\times$)
& 1.133 (2.30$\times$) \\

DB-BFS-Push
& 2.802 (1.11$\times$)
& 1.191 (1.00$\times$)
& 4.066 (1.31$\times$)
& 0.861 (1.69$\times$)
& 2.201 (1.74$\times$)
& 2.553 (1.59$\times$)
& 0.684 (1.97$\times$)
& 0.849 (1.73$\times$) \\

\textbf{DB-BFS-Auto}
& \textbf{2.526 (1.00$\times$)}
& \textbf{1.190 (1.00$\times$)}
& \textbf{3.095 (1.00$\times$)}
& \textbf{0.508 (1.00$\times$)}
& \textbf{1.262 (1.00$\times$)}
& \textbf{1.609 (1.00$\times$)}
& \textbf{0.347 (1.00$\times$)}
& \textbf{0.492 (1.00$\times$)}\\
\bottomrule
\end{tabular}
}
\end{table*}

We next evaluate whether the kernel-level advantages carry over to complete applications. 

\textbf{For BFS}, we run single-source BFS from vertex 0 for all methods to ensure a fair comparison. Although the source is fixed, the selected matrices cover diverse graph structures, including finite-element meshes, banded graphs, and irregular networks, leading to varied frontier evolution patterns that are representative of typical BFS workloads. This end-to-end measurement includes frontier generation, visited-state maintenance, path switching, and multi-iteration control overheads. On A100, DB-BFS achieves average speedups of 26.74$\times$, 13.36$\times$, 8.81$\times$, and 2.66$\times$ over Gunrock, Enterprise, GSwitch, and TileBFS, respectively. On RTX 4090, the corresponding speedups are 49.43$\times$, 12.40$\times$, 13.56$\times$, and 3.60$\times$. These gains come from representing frontiers directly at the block level, allowing sparse levels to avoid invalid edge scans through push and expanded levels to reduce writeback conflicts through pull.

\textbf{For sparse decoding}, we test Llama-3.2-1B\cite{grattafiori2024llama3herdmodels}, Qwen2.5-1.5B\cite{qwen2.5}, and Qwen3-0.6B\cite{qwen3technicalreport}. We apply Wanda\cite{sun2024simpleeffectivepruningapproach} to prune 50\% of the weights in each linear layer, while activation sparsity comes from magnitude-based thresholding\cite{liu2025trainingfreeactivationsparsitylarge}, yielding about 10\% activation density during single-token decoding. Tab.\ref{tab:decoding-comparision} reports the average per-token decoding time of the corresponding layers. DB-Decoding achieves consistent speedups across models and layers: Q/K/V/Out projections obtain about 2.35$\times$--2.50$\times$, while FFN projections show larger gains, up to 3.42$\times$ for Up and 4.50$\times$ for Down. Since weight matrices are reused after a single DB-format conversion and only activation masks change at runtime, DB-Decoding avoids repeatedly processing zero activations and zero-weight entries without maintaining multiple sparse weight layouts.

\subsection{Representative Analysis \& Ablations}

We use representative matrices to examine more closely when adaptive choices matter.

\begin{figure}[t!]
    \centering
    \includegraphics[width=1\linewidth]{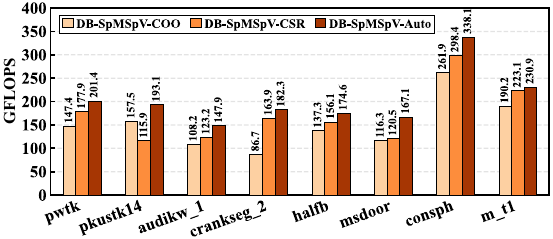}
    \vspace{-15pt}
    \caption{Ablation of low-level block storage formats in DB-SpMSpV. The input vector sparsity is 0.1, and all variants use the high-level pull computation path.}
    \vspace{-10pt}
    \label{fig:ablation-spmspv}
\end{figure}

For SpMSpV, we compare three low-level block storage strategies: DB-SpMSpV-COO, DB-SpMSpV-CSR, and DB-SpMSpV-Auto. Fig.\ref{fig:ablation-spmspv} shows that no fixed block format is consistently optimal across matrices. For example, \texttt{pkustk14} benefits more from the COO-like format, achieving 157.538 GFLOPS compared with 115.855 GFLOPS for the CSR-like format, while \texttt{crankseg\_2} favors the CSR-like format, reaching 163.85 GFLOPS compared with 86.75 GFLOPS for the COO-like format. DB-SpMSpV-Auto avoids this format-specific behavior and achieves the best performance on all representative matrices, reaching 201.353, 193.134, 182.02, and 338.129 GFLOPS on \texttt{pwtk}, \texttt{pkustk14}, \texttt{crankseg\_2}, and \texttt{consph}, respectively. In particular, Auto improves over COO-like by more than 2.0$\times$ on \texttt{crankseg\_2} and over CSR-like by about 1.67$\times$ on \texttt{pkustk14}. The contrast indicates that intra-block nonzero distribution and vector-block activity jointly shape the best local format, rather than either factor alone.

For BFS, Tab.\ref{tab:bfs-representative} reports representative matrices together with frontier evolution and end-to-end execution time. DB-BFS-Auto achieves the best performance on all representative matrices. On \texttt{audikw\_1}, where the frontier first expands and then shrinks, DB-BFS-Auto takes 3.095ms, outperforming Gunrock at 12.49ms, GSwitch at 66.33ms, and TileBFS at 3.49ms. On \texttt{halfb}, where the frontier is concentrated, and the matrix has a banded structure, DB-BFS-Auto reaches 1.262ms and still achieves a 2.94$\times$ speedup over TileBFS.

The single-path variants show the same pattern from the application side. On \texttt{halfb}, DB-BFS-Auto takes 1.262ms, outperforming DB-BFS-Push at 2.201ms and DB-BFS-Pull at 3.329ms. On \texttt{msdoor} and \texttt{consph}, Auto also outperforms the better single-path variant. BFS frontiers move through different sparsity regimes across iterations, so a fixed direction either wastes scans in sparse stages or increases writeback pressure in denser stages. DB-BFS avoids this by changing direction with the frontier state.

\subsection{Overhead}

Supporting both push and pull with full CSR+CSC duplication nearly doubles storage. By keeping only high-level dual views and reusing one low-level block payload, DB-SpMSpV saves 60.64\%--61.68\% storage over CSR+CSC in Tab.\ref{tab:memory_save}, with an average saving of 61.26\%. It also uses less memory than plain CSR on these matrices because low-level block formats reduce local indexing overhead: COO blocks pack local coordinates, Dense blocks remove per-element indices, ForceCSR/ForceCSC encode local structure with masks, and patched CSR uses compact row/mapping fields. Together, these designs reduce the amortized index footprint below the 32-bit column index used by standard CSR. The DB-format conversion is performed offline and reused across SpMSpV calls, BFS iterations, and decoding tokens; online overhead mainly comes from vector-mask generation, nonzero-block counting, path selection, and GPU kernel execution.

%%\vspace{-10pt}
\begin{table}[h]
\centering
\caption{Memory overhead comparison.}
\vspace{-5pt}
\label{tab:memory_save}
\begin{tabular}{ccccc}
\toprule
Method & pwtk & halfb & consph & m\_t1 \\
\midrule
CSR (MB)           & 133.98 & 142.62 &  69.10 & 111.99 \\
CSR+CSC (MB)       & 267.95 & 285.25 & 138.20 & 223.99 \\
\textbf{Ours (MB)} & \textbf{102.67} & \textbf{110.31} & \textbf{54.39} & \textbf{86.48} \\
Memory Save (\%)  & 61.68 & 61.33 & 60.64 & 61.39 \\
\bottomrule
\end{tabular}
\end{table}
%%\vspace{-10pt}

%% file: sections/related_work.tex
%\vspace{-15pt}
\section{Related Work}

\label{section:6}

Prior work has improved sparse matrix computation from three closely related directions: storage formats, SpMSpV kernels, and graph applications.

\textbf{Sparse blocked formats.} CSR/CSC remain the basic row- and column-oriented sparse formats, and many SpMV systems improve their locality, load balance, or format selection. CSR5 uses segmented tile descriptors for balanced SpMV, SMAT selects formats through input-adaptive tuning, and recent tiled/block formats improve GPU locality and regularity for SpMV-like kernels\cite{liu2015csr5,Li2013SMATAI,Niu2021TileSpMVAT,Cong2025CBSpMVADA,Greathouse2014CSR,du2022alphasparse}. TileSpMSpV is more directly related to our work: it extends tiling to GPU SpMSpV and also enables TileBFS by organizing matrices and vectors in tile form\cite{Ji2022TileSpMSpVAT}. Bit-GraphBLAS targets binary graph workloads with B2SR, a two-level representation that compacts block contents as bits\cite{Chen2022BitGraphBLASBO}. SpInfer uses bitmap encoding for GPU SpMM, while Sparse Register Tiling combines code transformation and data compression for CPU SpMM; both target static sparse weights with dense inputs rather than dynamic sparse vectors\cite{fan2025spinfer,wilkinson2023register}. These designs show the value of blocking and compact metadata. DB-SpMSpV differs in that it targets dynamic sparse vectors and general-valued SpMSpV: it keeps block-level CSR/CSC views only at the high level, reuses one low-level block payload, and still supports both push and pull without full CSR+CSC duplication.

\textbf{SpMSpV kernels and adaptivity.} Early SpMSpV work addressed irregular accumulation through graph-driven execution, k-way merge, sorting, and work-efficient parallel algorithms\cite{Yang2015FastSM,azad2017work,Li2018MergeBasedPS}. More recent methods focus on adaptivity and platform-specific bottlenecks: Adaptive SpMV/SpMSpV selects among multiple kernels according to sparsity and workload properties, fgSpMSpV recollects useful nonzeros for fine-grained execution, HAM-SpMSpV incorporates mask sparsity, and VDHA reduces GPU writeback conflicts with vector-driven hash aggregation and related optimizations\cite{Li2020AdaptiveSO,Chen2022fgSpMSpVAF,Xu2024HAMSpMSpVAO,Li2026VDHAVH}. These methods establish that input sparsity, load balance, and write conflicts dominate SpMSpV performance. Their adaptivity, however, is usually expressed as kernel selection or conflict handling on top of a given layout. DB-SpMSpV instead coordinates a shared-payload dual-view representation with global push/pull selection and local microkernel adaptation for dynamically changing sparse vectors.

\textbf{GraphBLAS and BFS applications.} GraphBLAS-style systems such as CombBLAS, GraphMat, GraphPad, and GraphBLAST formulate graph traversal through sparse linear algebra primitives\cite{Bulu2011TheCB,sundaram2015graphmat,anderson2016graphpad,yang2022graphblast}. Direction-optimizing BFS and push-pull GraphBLAS show that traversal direction should change as the frontier evolves, while GPU graph systems such as Gunrock and GSwitch expose frontier-centric scheduling and runtime pattern selection\cite{Beamer2012DirectionoptimizingBS,Yang2018ImplementingPE,Wang2017Gunrock,Meng2019APB}. CGA extends adaptive selection to masked SpMV/SpMSpV across heterogeneous platforms\cite{Xu2026CGAAB}. Specialized BFS accelerators further exploit tiled or binary structures: TileBFS builds on TileSpMSpV, BerryBees uses bit Tensor Cores, and BLEST improves TC-based BFS with a more balanced data structure and pipeline\cite{Ji2022TileSpMSpVAT,Niu2025BerryBeesBF,Elbek2025BLESTBE}. These systems confirm that BFS frontiers are dynamic and that fixed push or pull policies are insufficient. DB-SpMSpV follows this direction but makes the adaptation both reusable and fine-grained: the same SpMSpV abstraction supports DB-BFS and DB-Decoding, while path selection can occur across global traversal and local microkernels.

%% file: sections/conclusion.tex
\section{Conclusion}
\label{section:7}

This paper presents DB-SpMSpV, a dual-view blocked SpMSpV framework for dynamic GPU workloads. By separating high-level CSR/CSC traversal views from a shared low-level block payload, DB-SpMSpV supports push and pull without full format duplication. Its two-level adaptivity uses global nonzero-block statistics and local vector-block distribution to choose traversal paths and microkernels, while load balancing, asynchronous prefetching, and hierarchical writeback address GPU sparse-computation bottlenecks. Experiments on A100 and RTX 4090 show that these methods improve kernel-level SpMSpV and accelerate end-to-end DB-BFS and single-token linear layers in DB-Decoding.

\section*{Limitations}
DB-SpMSpV is designed for sparse input vectors; extending it to SpMM would require different data reuse and accumulation strategies, while conventional SpMV may be preferable for dense inputs. DB-Decoding is evaluated only at the sparse linear-layer level, so its end-to-end impact depends on how much decoding time these layers account for. Our evaluation is limited to A100 and RTX 4090. The irregular execution pattern does not map directly to Tensor Cores or TMA, whose use would require specialized layouts and kernels. Full-model evaluation and these hardware-specific extensions remain future work.